\documentstyle[12pt]{article}
\begin{document}
\bibliographystyle{plain}
\setlength{\baselineskip}{.7cm}
\renewcommand{\thefootnote}{\fnsymbol{footnote}}
\sloppy
 
\vspace*{2.0in}
 
\begin{center}
\centering{\Large\bf The Paradox of the expected Time until the Next
Earthquake}
\end{center}
 
\vspace {1.0in}
 
\begin{center}
\centering{\large\bf D. Sornette and L. Knopoff}
\end{center}
 
\vspace {0.25cm}
 
\begin{center}
\centering{\bf Institute of Geophysics and Planetary Physics\\
University of California, Los Angeles}
\end{center}
 
\renewcommand{\thefootnote}{\fnsymbol{footnote}}
 
\newpage
 
\noindent
{\large\bf Abstract}

We show analytically that the answer
to the question, ``The longer it has been since the last
earthquake, the longer the expected time till the next?''
depends crucially on the statistics of the
fluctuations in the interval times between earthquakes. The periodic,
uniform, semigaussian, Rayleigh and truncated statistical
distributions of interval times, as well as the Weibull distributions
with exponent greater than 1, all have decreasing expected time to the
next earthquake with increasing time since the last one, for long
times since the last earthquake; the log-normal and power-law
distributions and the Weibull distributions with exponents smaller
than 1, have increasing times to the next earthquake as the elapsed
time since the last increases, for long elapsed times.  There is an
identifiable crossover between these models, which is gauged by the
rate of fall-off of the long-term tail of the distribution in
comparison with an exponential fall-off. The response to the question
for short elapsed times is also evaluated. The lognormal and power-law
distributions give one response for short elapsed times and the
opposite for long elapsed times.
Even the sampling of a finite number of intervals from
the Poisson distribution will lead to an
increasing estimate of time to the next earthquake for increasing
elapsed time since the last one.
 
\pagebreak
 
\begin{center}
\centering{\large\bf Introduction}
\end{center}
 
While small earthquakes after removal of aftershocks have a Poissonian
distribution (Gardner and Knopoff, 1974),
intermediate and large earthquakes in a given region are clustered in
time (Kagan and Knopoff, 1976; Lee and Brillinger, 1979; Vere-Jones
and Ozaki, 1982; Grant and Sieh, 1994; Kagan and Jackson, 1991, 1994;
Kagan, 1983; Knopoff, et al., 1996); `clustering' is taken to mean
that the earthquakes do not have a purely Poissonian, memoryless
distribution of time intervals. According to this interpretation,
periodic earthquakes are the extreme limit of clustering and can be
predicted exactly.   More generally, if there is temporal clustering,
the estimate of the probability of occurrence of a future earthquake
in a given time interval is improved if there is a knowledge of the
times of previous events, since clustering implies a memory. To
express this property quantitatively, we relate the elapsed time since
the last earthquake in a region  to the conditional probability of
occurrence of the next earthquake within a given time interval from
the present. Davis, et al. (1989) have posed a version of this problem
in the form of the question,
\vspace{0.1in}
 
\noindent
{\bf (Q.)}~~~~~~~{\it (can it be that) ``The longer it has been since
the last earthquake, the longer the expected time till the next?"}
 
\vspace{0.1in}
\noindent
The observation of Davis, et al. for the log-normal distribution was
that the answer to Q. is positive. Ward and Goes (1993) and Goes and
Ward (1994) showed numerically that, in the case of the Weibull
distribution, the response to Q. can be either yes or no, depending on
the exponent in the distribution. The positive responses would seem to
be counterintuitive, since it is to be expected that an earthquake
should be more likely to occur with increasing time in response to an
inexorable tectonic loading that brings a fault ever closer to its
finite threshold of fracture.
 
The intuitive interpretation is of course consistent with simple
relaxation oscillator models of the earthquake process, such as the
slip- or time-predictable models.  But these models should be
reconsidered if the stress field is altered on a given fault segment
due to redistribution derived from earthquakes on nearby fault
elements; these interactions can cause fluctuations in the stress
field, with consequent fluctuations in the interval times.  Knopoff
(1996) has proposed that the fluctuations in the interval times
between great earthquakes on the San Andreas Fault (Sieh, et al.,
1989) may be associated with stress interactions between the San
Andreas Fault and other, nearby faults.
 
Below, we give a rigorous statistical framework for the derivation of
a quantitative response to Q. Statistical estimates of recurrence
times will be found to be very sensitive to assumptions about
statistical distributions. Our results confirm, quantify and extend
the numerical analyses of Davis, et al. (1989) and Ward and Goes
(1993) and Goes and Ward (1994), by providing an analytic basis for
the problem.
 
\vspace {0.5cm}
 
\begin{center}
{\large\bf The time to the next earthquake}
\end{center}
 
\vspace{0.1in}
 
Let $p(t)$ be the probability density of the time intervals between
earthquakes. If the time (now) since the last earthquake is $t$, what
is the probability density function $P(t')$ that we must wait an
additional time $t'$ until the next earthquake? From Bayes' theorem
for conditional probabilities, cited in elementary statistics
textbooks, the probability that an event $A$, given the knowledge of
an event $B$, is simply the quotient of the probability of the event
$A$ without constraint and the probability of event $B$:
$$ P(A | B) = {P(A) \over P(B)}   .   \eqno(1) $$
Applied to this problem, $P(A) = p(t+t')$ which is the probability
that the next earthquake will occur at time $t'$ from now, and $P(B) =
\int_t^{\infty} p(s) ds$, which is the probability that no earthquake
has occurred up to now. Thus
$$ P(t') = {p(t+t') \over \int_t^{\infty} p(s) ds}  ,  \eqno(2) $$
which is normalized.
 
We calculate the expected time until the next earthquake $\langle t'
\rangle$ as a function of the time since the last one.
\vspace{0.1in}
 
\noindent
{\bf (A.)} ~~~~~~~{\it The answer to Q. is given by the sign of
${d{\langle t' \rangle} \over {dt}}$, if $\langle t \rangle$
exists.}
\vspace{0.1in}
 
\noindent
From eq. (2), the average expected time to the next earthquake is
$${\langle t' \rangle} ={ {\int_0^{\infty} t'p(t+t')dt'}
\over {\int_t^{\infty} p(u) du}}  .  \eqno(3)  $$
By a simple change of variable,
$$  {\langle t' \rangle}= {{\int_t^{\infty} (u - t)p(u)du }
\over {\int_t^{\infty} p(u) du}}  .  \eqno(4) $$
We integrate the numerator of (4) by parts and get
 
$$  {\langle t' \rangle}= {{\int_t^{\infty} ds \int_s^{\infty} p(u)du }
\over {\int_t^{\infty} p(u) du}}  .  \eqno(5) $$
The  denominator and numerator of (5) are the familiar first
cumulative integral and the less familiar second cumulative integral
of $p(u)$. For simplicity, we write (5) as
$${\langle t' \rangle} = -{ {f(t)} \over {f'(t)}} \eqno(6)$$
where  $f''(t) = p(t)$, i.e. $f(t)$
is the second cumulative integral of $p(u)$.
Thus $${{d{\langle t' \rangle}} \over {dt}} > 0~~~~~if~~~f(t)f''(t) -
[f'(t)]^2  >0~~~. \eqno(7a)$$
Equivalently,
$${{d{\langle t' \rangle}} \over {dt}} = - {g''(t) \over [g'(t)]^2} >
0~~~~~if~~~{g''(t) < 0}~~~~~~. \eqno(7b),$$
where we have set $f(t) = e^{-g(t)}$. The signs are appropriately
reversed in the case $g''(t) > 0$. Equation (7b) especially favors an
appreciation of the behavior at large values of elapsed time $t$.
 
If $p(t)$ is finite at $t=0$, we can find yet a third version of (5)
which is useful for  $t$ small. A straightforward expansion for small
$t$ shows that
$$\lim_{t\rightarrow 0^+}{{d{\langle t' \rangle}} \over {dt}}  =
p(0)\Delta - 1 ~~~~,
\eqno(7c) $$
where
$$\Delta = \int_0^{\infty} ds \int_s^{\infty} p(u)du =
\int_0^{\infty} s p(s)ds \equiv \langle t \rangle ~~~~.
$$
The second integral on the r.h.s. is obtained by integration by parts;
$\langle t \rangle$ is the average (unconditional) time of recurrence
between two earthquakes. Let $\tau \equiv {1 \over p(0)}$, where
$\tau$ is the estimate of the waiting time until the next earthquake
made immediately after the occurrence of the preceding earthquake. We
call $\tau$ the instantaneous estimate of $t'$. Thus, ${{d{\langle t'
\rangle}} \over {dt}}$ can be rewritten in the simple form
$$
\lim_{t\rightarrow 0^+}{{d{\langle t' \rangle}} \over {dt}}  =
{\langle t \rangle \over \tau} - 1  ~~~~.
\eqno(7d)
$$
\begin{itemize}
\item If the instantaneous estimate $\tau$ of the waiting time is
smaller than the average waiting time $\langle t \rangle$, the time to
the next earthquake increases with increasing time since the last one
for small $t$\,: this reflects the fact that the average waiting time
$\langle t \rangle$ is formed by contributions from the distribution
over all time, and a value of $\langle t \rangle$ larger than $\tau$
indicates contributions from the distribution that are larger than
$\tau$ at non-zero times; in this case ~~~$ \lim_{t\rightarrow
0_+}{{d{\langle t' \rangle}} \over {dt}} > 0$. \item If  $\langle t
\rangle < \tau$, the reverse is true, shorter and shorter time scales
are sampled on the average as time increases, and the time to the next
earthquake decreases with increasing time since the last one for small
$t$.
\end{itemize}
In particular, if $p(0)=0$, then ~~~$\lim_{t\rightarrow
0_+}{{d{\langle t' \rangle}} \over {dt}} = -1$, and the time to the
next earthquake decreases with increasing time the since last one for
small $t$; if however $p(0)=\infty$, then ~~~$\lim_{t\rightarrow
0_+}{{d{\langle t' \rangle}} \over {dt}} = \infty,$ and the time to
the next earthquake increases with increasing time since the last one
for small $t$.
 
The generalization of (7c) to times other than $t=0$ is the criterion
(7a), when the mean $\langle t \rangle$ exists. When it does not
exist, as for example when the tails of the distributions decay slower
than $t^{-2}$, we must compare $P(t')$ as given by (2), with $p(t')$.
 
\vspace{0.5cm}
 
\begin{center}
\centering {\large\bf Exponential distribution}
\end{center}

\vspace{0.1in}
 
In order to develop some intuition, we first consider the exponential
distribution, which is the familiar case of Poissonian statistics,
$$ p(t) ={{e^{-t/t_0}}\over {t_0}}  ,  ~~~~~~~~~t \geq 0  , $$
where $t_0$ is the mean interval time between earthquakes. From eq.
(2),
$$  P(t') = {{e^{-t'/t_0}}\over {t_0}}.   \eqno(8) $$
Not unexpectedly, the estimate of the time of occurrence of the next
earthquake does not depend on the elapsed time: the average time from
now to the future earthquake is $t_0$, no matter what the value of
$t$. This case is memoryless; indeed it is the only distribution that
has no memory. The expected time to the next earthquake is ${\langle
t' \rangle} = t_0$; there is no need to invoke the machinery of (7) to
derive ${d{\langle t' \rangle} \over {dt}} = 0$. The Poisson
distribution is the unique case $g''(t) = 0$, which gives the same
result.
 
The exponential distribution is the fixed point of the transformation
$p(t) \rightarrow P(t')$,  {\it i.e.} it is the solution to the
functional equation $$P(t')=p(t').  \eqno(9)$$ To verify that (8) is
the solution to (9), differentiate (2) with respect to $t$, and
substitute in  (9).  We get
$$ -p(t)p(t') = {{dp(t+t')} \over {dt}}.  \eqno(10) $$
Take the Laplace transform of (10) with respect to $t'$, with $1/t_0$
the transform variable.  The result (8) follows. Thus the exponential
distribution is the fixed point of (2).
 
We restate these results: Except for the Poisson distribution, all
statistical distributions must have an average time from now to the
future earthquake that depends on the time since the last earthquake.
If the long-time tail of the function $f(t)$, defined as the integral
of the integral of the distribution $p(t)$, falls off at a rate that
is faster than exponential, the expected time to the next earthquake
is reduced, the longer the elapsed time since the last, and vice
versa.  The Poisson distribution is the cross-over between the two
states. The exponential case has neither a positive nor a negative
response to Q., since the time since the last earthquake has no
influence on the time of the next.
 
\vspace {0.5cm}
 
\begin{center}
\centering {\large\bf Other Conditional Distributions}
\end{center}
 
We calculate the expected time to the next earthquake for several
examples of statistical distributions $p(t)$ with memory. We
illustrate the results in figure 1 by displaying the average time to
the future earthquake $t'$ plotted against the time since the last
earthquake $t$ for selected distributions $p(t)$;  the positive or
negative slopes of the curves give the answers to question Q. The
details of the calculations are given in the Appendix.
 
The analytical results are summarized in Table 1.
The times in the table are scaled by a characteristic time $t_0$ for a
given distribution; the precise definition of $t_0$ for each
distribution is given in the Appendix. In general, $t_0$ is of the
order of the mean time between earthquakes, if it is not so exactly.
In most cases we can give an answer A. that is valid over the entire
range of elapsed times since the last earthquake.  In some cases, we
can only give the answer for the limits of short and long time since
the last earthquake.  The table is arranged to favor the limiting
responses, even though we may have the complete solution. In the case
of the exponential distribution, the response is neutral as we have
already discussed.  In the cases of the periodic and uniform
distributions, the response is only meaningful for times up to $t_0$;
in the case of the Gaussian distribution, the response is meaningful
for long elapsed times, as we discuss in the appendix. In the cases of
the Weibull distribution with $m > 1$ and the semigaussian
distribution, the response can be proved to be negative for both short
and long elapsed times, and can be inferred to be negative for all
elapsed times.  For the Weibull distribution with $m < 1$ and the
power-law distribution, the response is positive for both short and
long elapsed times, and can be inferred to be positive for all elapsed
times.  For the lognormal, power-law and truncated power-law
distributions there is one response for short elapsed times and the
opposite response for long elapsed times, with the implications of a
crossover and hence neutral response at an intervening time scaled by
$t_0$; the lognormal and truncated power-law distributions have
opposite responses to each other in the long and short time regimes.
 
Since the truncated and ordinary power-law distributions give opposite
results for long elapsed times, it follows that the answers to Q. are
unstable with respect to the presence or absence of a cutoff in the
distributions.  It is by definition problematical that a presumed
existence of a cutoff can be identified from a finite set of
observations of interval times: there is no guarantee that a presumed
cutoff will not disappear with a future observation of a longer
interval between earthquakes.  Thus a positive response to the
question for long elapsed times since the last earthquake is only
conjectural, i.e. it is only as strong as one knows the distribution
to times longer than have been observed, which is impossible.  Of
course, the distribution can always be postulated {\it a priori}, as
in the numerical examples of Davis, et al. (1989)  and Ward and Goes
(1993), but the postulate does not ensure that it represents nature.
 
\vspace {0.5cm}

\begin{center}
\centering{\large\bf Estimate of $t_0$}
\end{center}
 
Suppose that we do not know {\it a priori} the characteristic time
$t_0$ of time intervals between successive earthquakes. We then have
to estimate it from a finite suite of observations of interearthquake
time intervals. Assume that $(n - 1)$ observations of time intervals
$t_1$, $t_2$, ... , $t_{n-1}$, are made precisely;  we ignore here the
additional problem of the uncertainties in the time intervals that
occur for historical earthquakes; this can be treated by standard
statistical methods (Sieh, et al., 1989). Suppose that the time since
the last event is $t$. Then, in the case of the Poisson distribution
$p(t) ={{e^{-t/t_0}}\over {t_0}}$, the standard maximum likelihood
method gives the estimate of $t_0$ as the value which maximizes ${1
\over t_0^n} e^{-{t + \sum_{j=1}^{n-1} t_j \over t_0}}$, {\it i.e.}
$$ t_0 = {1 \over n} \biggl( t + \sum_{j=1}^{n-1} t_j \biggl)  .
\eqno(11) $$
Thus, even for the Poissonian case, the use of the information that no
event has occurred since $t$, gives an estimate of the average
recurrence time $t_0$ for the next event that increases with $t$\,!
The Poisson distribution is memoryless only if its parameter $t_0$ is
known {\it a priori}.
 
The calculation (11) can be generalized for the other distributions
discussed above. Our previous results must thus be reconsidered if the
parameters of the distributions are themselves not known precisely but
are estimated using presently available information. This does not
pose any difficulty in principle but must be addressed case by case.
This simple calculation demonstrates the sensitivity of the
``prediction'' to the assumptions concerning what is really known and
what is only inferred from the data.
 
\vspace{0.5cm}
 
\begin{center}
\centering{\large\bf Summary}
\end{center}
 
These observations can be summarized as follows:
\begin{itemize}
\item The Poisson or exponential distribution is memoryless and the
expected time until the next event is independent of previous
observations and of the elapsed time since the last earthquake. The
exponential thus acts as a fixed point in the space of distributions
of the transformation (2), and sits at the boundary between the
positive and negative classes of memory, i.e. at the boundary between
positive and negative responses to Q. Any statistics of the
fluctuations of recurrence times that is different from Poissonian
entails the explicit assumption of a memory. \item  {\bf a)} Any
distribution that falls off at large time intervals at a faster rate
than an exponential, such as the periodic, quasiperiodic, uniform and
semigaussian distributions and the Weibull distribution with $m>1$,
has the property, ``the longer it has been since the last earthquake,
the shorter the expected time until the next''.  The truncated power-law
distribution for times close to the cutoff time also has this
property.
 
\noindent
{\bf b)} Any distribution that falls off at large time intervals at a
slower rate than an exponential, such as the Weibull distribution with
$m<1$, the unbounded lognormal and power-law distributions, and the
truncated versions of these distributions for times remote from the
cutoff,  have the property, ``the longer it has been since the last
earthquake, the longer the expected time till the next''.
 
\item {\bf a)} All distributions that have an instantaneous
expectation time interval between earthquakes {\it smaller} than the
average waiting time between earthquakes have the property of an {\it
increasing} time to the next earthquake for an increasing time the
since the last one, for short times since the last one.  This includes
the cases $p(0) = \infty.$
 
\noindent
{\bf b)} All distributions that have an instantaneous expectation time
interval time between earthquakes {\it larger} than the average
waiting time between earthquakes have the property of a {\it
decreasing} time to the next earthquake for an increasing time since
the last one, for short times since the last one.  This includes the
cases $p(0) = 0.$
 
\item Caution should be exercised in the use of
statistics of fluctuations of interval
times deduced from  data sets that describe only the distributions for
short time intervals between earthquakes. This is because of the
strong dependence of our result for long times, and in some cases for
short times as well, on the properties of the tails of the
distributions as well as on the values of the parameters of the
distributions.
 
\item The estimate of the time until the next earthquake depends on a
precise estimate of the tail of $p(t)$ and is unstable with respect to
presently available data for the recurrence of large earthquakes.
Even a finite sampling of the Poisson distribution will lead to an
estimate of the time to the next earthquake that increases with
increased time since the last one.
 
\end{itemize}
 
Thus the positive response of Davis, et al. (1989) to Q., ``the longer
it has been since the last earthquake, the longer the expected time
till the next'', is shown to arise from the use of a distribution that
decays slower than an exponential and which is unbounded for large
time intervals; the result is valid for other distributions as well.
If a slowly decaying law itself undergoes a transition at even longer
intervals to a more rapidly decaying law, as in the extreme case of a
distribution with a cutoff, one can expect that eventually the next
earthquake will become more and more probable. The response to the
question (Q.) is also related, in part, to the finiteness of the
number of observations of time intervals between earthquakes that give
the estimate of the distribution $p(t)$; the extrapolation of the
estimate of the distribution to its asymptote for very large time
intervals is exceedingly dangerous, since this is procedure is likely
to be based on few, if any, observations. The results of this exercise
suggest that caution be used in the extrapolation of statistics
deduced from short time-scale data sets to long time-scales.
\newpage
\begin{center}
{\large\bf Appendix}
\end{center}
 
\vspace {0.5cm}
We calculate the expected time to the next earthquake for several
examples of statistical distributions $p(t)$ with memory.
 
\vspace {0.5cm}
\noindent
{\large\bf A. Periodic distribution}
 
The simplest of the distributions with memory is the periodic
distribution,
$$p(t) = \delta (t_0 - t) .$$ By inspection, we have
$$P(t') = \delta (t_0 - t' - t) , ~~~~~~{\langle t' \rangle} = t_0 - t .$$
Without invoking the generalized machinery, ${{d{\langle t' \rangle}}
\over {dt}} = -1$.  In this simplest of cases, the expected time of
the forthcoming earthquake decreases as the elapsed time since the
preceding earthquake increases.  Extension to quasiperiodic cases can
be made.
\vspace {0.5cm}
 
\noindent
{\large\bf B. Uniform distribution}
\vspace {0.1in}
 
The uniform distribution is
$$ p(t) = {1 \over t_0}   ~~~~~~~~~0 \leq t \leq t_0 , $$
where $t_0$ is the maximum interval between earthquakes and $t_0/2$ is
the mean. From eq. (2), we get
$$ P(t') =  {1 \over t_0-t}  ~~~~~~~~~~~0 \leq t' \leq t_0-t .
\eqno(A1) $$
$P(t')$ is independent of $t'$, {\it i.e.} it is itself a uniform
distribution, but its value is dependent on $t$. The probability that
an earthquake will occur at any time in the future up to $t_0$
increases as the time since the last earthquake increases, and becomes
infinite as $t \rightarrow t_0$, which simply expresses the intuitive
result that the event will occur with certainty before $t_0$. It is
easy to see that the average time to the future earthquake from the
present is $\langle t' \rangle = {1 \over 2}(t_0 - t). $ The negative
value of $d\langle t'\rangle /dt $ gives the answer (A.): the expected
time to the next earthquake decreases with increasing time since the
last earthquake. In figure 1, we show the average time to the future
earthquake plotted against the time since the last earthquake; both
coordinates are normalized by the mean time between earthquakes,
$t_0/2.$ The linear relationship between $\langle t' \rangle$ and $t$
is strongly curved on the log-log plot.
 
In Fig. 2, we display the probability that the next earthquake will
occur at time $t'$ from now. We show the unconditional probability,
i.e., the probability as though we knew the distribution of intervals
$p(t)$ but did not know the time of the last earthquake.  We also show
the (conditional) probability of an earthquake in the future knowing
that the last earthquake took place at time $1.33 t_0$ in the past.
In the latter case, no earthquake can occur after $0.67t_0$ from now;
according to (A1), the probability of occurrence of the future
earthquake is higher by a factor of 3 than the unconditional
probability, and is independent of the time of the future earthquake.
 
\vspace {0.5cm}
 
\noindent
{\large\bf C. (Semi)-Gaussian distribution}
\vspace {0.1in}
 
The Gaussian distribution is
$$p(t) = {1 \over {\sqrt {2\pi} \sigma}} e^{-{(t-t_0)^2 \over 2\sigma^2}}~~~,$$
where $t_0$ is both the mean and the most probable time interval of
earthquake recurrence; the standard deviation is $\sigma$. The
Gaussian distribution has a finite probability that the next
earthquake will occur before the preceding one. The drawback is minor
if $\sigma \ll t_0$, a condition that describes a nearly periodic
distribution; we have considered the periodic case above.  To restrict
the problem to cases of positive $t$, we could truncate the
distribution at $t=0$, however this leads to messy mathematics; for
large times, the drawback is minor.
 
For the simpler problem $t\gg t_0$, we use the approximation
$$p(t) = {2 \over {\pi t_0}} e^{-{1 \over \pi}({t \over t_0})^2},
~~~t \geq 0$$
which is the semi-gaussian distribution, i.e. it is a Gaussian
centered at $t=0$, the mean time interval of earthquake recurrence is
$t_0$ and the most probable time for recurrence is of course zero for
this distribution. Formally, eq. (2) yields
$$ P(t') = {2 \over \pi t_0} {e^{-{(t+t')^2 \over \pi t_0^2}}
\over {\em erfc}({t \over \pi t_0})}~~~~~~~~~, \eqno(A2) $$
where ${\em erfc}(x)$ is the usual complementary error function.
 
If the elapsed time since the last earthquake is very large, $t \gg
t_0$, we can use the first term of the asymptotic expansion of the
second cumulative integral which is $f(t)\sim {e^{-{t^2 \over \pi
t_0^2}} \over {t^2}}$. Then $g(t) \sim t^2 + {O} (\log t),$ $g''(t) >
0$ for large $t$, and from (7b)
$${{d{\langle t' \rangle}} \over {dt}} < 0~~~~~~~~~. $$
For short times, we use (7c) with $p(0)= {2 \over \pi t_0}$ and
$\Delta = t_0$ and get
$${{{d{\langle t' \rangle}} \over {dt}} \approx {{2}\over {\pi}} - 1 =
-0.363}.$$
Thus in both the short and long time limits, the expected time until
the next earthquake decreases as the time since the last one
increases.

\vspace {0.5cm}
\noindent
{\large\bf D. Lognormal distribution}
\vspace {0.1in}
 
The lognormal distribution is
$$
p(t) = {1 \over \sqrt{2 \pi} \sigma} {1 \over t}
e^{-{(\log{t \over t_0})^2 \over 2 \sigma^2}} =
{1 \over \sqrt{2 \pi} \sigma t_0} e^{-{1 \over {2 \sigma^2} } (\log{t \over t_0}
+ \sigma ^2)^2 + {\sigma ^2 \over 2}}
.  \eqno(A3)$$
which is similar in shape to the Rayleigh distribution (see below)
near $t=0$ but has a much more slowly decaying tail for large times.
In this case, $t_0$ is the median time; the mean time is $\langle t
\rangle = t_0 e^{\sigma^2 \over 2}$; the most probable time is $t_0
e^{-\sigma^2}$. From (A3), we can write
$$ g(t) = (\log({t\over t_0}) + \sigma ^2)^2 + {O}(\log \log ({t\over
t_0})), $$
whence
$$g''(t) \approx {2\over t^2}(1 - \sigma ^2 - \log {t\over t_0}).$$
For $ t \gg t_0, ~~~g''<0$ and hence ${{d{\langle t' \rangle}} \over
{dt}} > 0.$  For $t \ll t_0,  ~~~g''>0$ and ${{d{\langle t' \rangle}}
\over {dt}} < 0;$ alternatively, we note that $p(0) = 0$ and hence
from (7c), ${{d{\langle t' \rangle}} \over {dt}} < 0,$ which is the
same result.   Thus the lognormal distribution has a crossover in
response to Q.
 
We express these results graphically. For the case $\sigma = t_0$ we
display $P(t')$ for times $t=2t_0$ and $t=5 t_0$ (Fig. 3) together
with the unconditional log-normal distribution $p(t)$. From figure 3a,
we see that $P(t')$ is significantly smaller than $p(t')$ for times
comparable to the elapsed time $t$ but $P(t')$ is, as expected, larger
than $p(t')$ at large times (see extension of Fig. 3a to long times in
Fig. 3b). This is a small effect for $t=2t_0$ but is much stronger for
$t=5 t_0$ and all the more so if $t$ increases even more. Thus,
numerically as well as analytically,  for early elapsed times $t$ that
are comparable to the peak of the distribution, the longer the elapsed
time since the last event, the shorter the time until the next event;
but for large elapsed times since the last earthquake, the longer the
time since the last event, the longer the time until the next one. In
the lognormal case, it is correct to state that ``the longer it has
been since the last earthquake, the longer the expected time until the
next'' but only for elapsed times greater than times of the order of
the characteristic time.  The lognormal distribution is an example of
a case that has one answer to Q. for short times since the last
earthquake and the opposite answer for long times. Of course, the
crossover takes place at elapsed times that are of the order of the
characteristic time $t_0$. Note that the probabilities for long times
in the future are, as might be expected, very small.  The log-normal
case  is an example of a distribution with a tail that decays slower
than an exponential.
 
To illustrate more concretely the properties of a system that has a
crossover response to Q., we concoct the distribution
$$p(t) = {1\over {4t_0}} {\sqrt {t\over t_0}} e^{-{\sqrt {t\over t_0}}}$$
which probably has no redeeming virtue in nature, but has the property
that it has easily calculable integrals.  It is evident that this
distribution has both a long-time tail that decays slower than
exponential, and the property $p(0)=0$, as in the case of lognormal
distribution. Thus we are guaranteed that there is a crossover in the
response to Q. More precisely, the criterion function (7a) is
$f(t)f''(t) - [f'(t)]^2  = {1 \over 4} (x^3+4x^2+4x-4) e^{-2x}$\,
where $x= {\sqrt {t\over t_0}}.$ The criterion has the crossover in
sign between $x$ small and large at ${t\over t_0}=0.3532$.  The
lognormal distribution has similar properties.
 
\newpage
 
\noindent
{\large\bf E. Weibull distribution}
\vspace {0.1in}
 
The Weibull distribution is
$$ p(t) = m t_0^{-m} t^{m-1} e^{-({t \over t_0})^m} , ~~~~~~~~~ 0 < t <
\infty, ~~~~~~~~~~~~~~m > 0 $$
having a most probable value $(m-1)^{1 \over m}t_0 $, a mean
$\alpha(m) t_0$ where $\alpha(m) = \int_0^{\infty} e^{-t^m} dt,$ and
median $(\log 2)^{1/m} t_0 $. Values of $m$ smaller than $1$
correspond to $p(t)$ decaying slower than an exponential for large $t$
and give the so-called stretched exponential distributions, while
values of $m$ larger than $1$ lead to a decay that is faster than
exponential. For large $m$, $p(t)$ approaches a delta function
centered on $t_0$, {\it i.e.} to the periodic distribution we have
considered above. Ward and Goes (1993) and Goes and Ward (1994) have
studied the degree of earthquake clustering as a function of $m$; in
their notation, $\nu = 1/m$. Equation (2) yields
$$ P(t') = m t_0^{-m} (t+t')^{m-1} e^{-{(t+t')^m - t^m \over t_0^m}}
. \eqno(A4) $$
The first term of the asymptotic series for $f(t)$, which is the
second cumulative integral  of $p(t)$, is
$$ f(t) \sim {e^{-({t \over t_0})^m}\over {({t \over t_0})^{3(m-1)}}} =
{e^{-({t \over t_0})^m - (3m-1)log{t \over t_0}}}. $$
Evidently $g(t) \sim t^m + O(log t)$ and hence $g''(t) \sim
m(m-1)t^{(m-2)}$. Thus if $0<m<1$, then $g''(t) < 0$, and ${{d{\langle
t' \rangle}} \over {dt}} > 0 $ for large $t,$ while if $m>1$, then
$g''(t) > 0$, and ${{d{\langle t' \rangle}} \over {dt}} < 0 $ for
large $t.$ For small $t,$  $p(0)=0, m> 1$ and it follows from (7c)
that ${d{\langle t' \rangle} \over {dt}} < 0.$  For cases $m < 1$,
$p(0)=\infty$, and from (7c), ${d{\langle t' \rangle} \over {dt}} >
0.$ Thus there is a reversal in response between the cases $m < 1$ and
$m > 1$, in agreement with the result of Ward and Goes.
 
In figure 4a, we exhibit the interesting subcase of the Weibull
distribution  with $m=2$ which is appropriate for rectified Gaussian
noise and is known as the Rayleigh distribution. The Rayleigh
distribution has a tail with similar properties to that of the
Gaussian, and decays faster than an exponential. We show $P(t')$ for
times $t=t_0$ and $t=2t_0$. It is clear that $P(t')$ has a
progressively shrinking width to the origin as $t$ increases, {\it
i.e.} the expected  time decreases as the waiting time $t$ increases.
The answer A. is negative for both short times and long times, the
latter property evidently connected with the rapid fall-off in $p(t)$
for large $t$.
 
The opposite situation is found in the case $m < 1$; in figure 4b we
plot the case $m=1/2$ and show  $P(t')$ for times $t=t_0$, $t=2 t_0$
and $t=10 t_0$, as well as $p(t)$. It is clear that  $P(t')$ lies well
above $p(t')$ at long times $t' > t_0$, and all the more so as $t$
increases. Thus the longer we wait, the longer the time to the next
event, in this case.
 
\vspace {0.5cm}
 
\noindent
{\large\bf F. Power-law distribution}
\vspace {0.1in}
 
The unconditional power-law distribution is
$$p(t) = 0, \hspace*{0.75in} ~~~~~~~~~~~ 0 <t <t_0, \hspace*{1.4in}
\eqno(A5)$$
$$ p(t) = {\mu \over t_0} \biggl({t \over t_0}\biggl)^{-(1 + \mu)} ~~~~~~~~
t_0 \leq t \leq \infty, ~~~~~~~~~~~ 0 < \mu < \infty$$
The characteristic time scale $t_0$ is proportional to the mean
${{\mu} \over {\mu - 1}}t_0$ for $\mu > 1$ and the median $t_0 2^{1
\over \mu}$; for $\mu < 1$, the mean is infinite. This is an example
of a distribution with a waiting time and has been used in the case
$\mu = 1/2$ in short-term earthquake prediction calculations by Kagan
and Knopoff (1981, 1987).
 
For $\mu > 1$, we evaluate ${d{\langle t' \rangle} \over {dt}}$ for
this case by applying the criterion function (7a), which gives $\left(
{t_0}\over {t} \right)^{2\mu} {1\over {\mu - 1}}.$  Thus ${d{\langle
t' \rangle} \over {dt}} > 0 $ for all $\mu > 1$ and all $t > t_0$.
 
For $\mu \leq 1$, $\langle t \rangle$ diverges and the criterion (7a)
cannot be used. In this case, we have to examine the conditional
distribution (2) directly, and compare it with the unconditional
distribution. This general method is also applicable to the case $\mu
> 1$.  Substitution of (A5) in eq. (2) yields
$$ P(t') =  {\mu \over t} (1 + {t' \over t})^{-(1 + \mu)}  . \eqno(A6) $$
Formula (A6) is almost the same as the unconditional distribution,
except for the additional $1$ in the parentheses; the two expressions
are identical in the limit $t' \gg t$ except for obvious scaling
factors. Thus in this limit, the distribution $P(t')$ is also a power law
with a characteristic scale given by the waiting time $t$, instead
of $t_0$ for the unconditioned $p(t')$. Thus  the longer it has been
since the last earthquake, the longer the expected time until the
next, for all cases $\mu > 1$.
 
Figure 5a shows $P(t')$ for $t=10 t_0$ and $t=100 t_0$ together with
$p(t)$ for an exponent $\mu = 3$ (for this choice, $p(t)$ possesses a
finite mean and variance). We observe the asymptotic power-law
behavior of $P(t')$ at times $t'>t$ with amplitude significantly
larger than $p(t')$, showing the enhanced probability for large
conditional waiting times. Figure 5b shows $P(t')$ for $t=10 t_0$ and
$t=100 t_0$ together with $p(t)$ for the threshold case of exponent
$\mu = 1$; in this case, the mean and variance are not defined. This
is an illustration of a power law with a very weighty tail. The
behavior of $P(t')$ is qualitatively similar to the previous case.
 
\vspace {0.5cm}
 
\noindent
{\large\bf G. Truncated power-law distribution}
\vspace {0.1in}
 
Except for the uniform and periodic distributions, we have considered
thus far only distributions of fluctuations in interval times that
extend to infinity.  In these cases of distributions with long-time
tails, there is a finite but small probability that a second
earthquake will occur after a very long time interval after the first.
If the distributions describe the seismicity of a region, rather than
that of an individual fault, the very long time intervals imply very
large accumulations of deformational energy and hence very large
fracture sizes.  To avoid the problems of earthquake sizes greater
than the size of a given region, we consider a cutoff in the
distributions $p(t)$ (Knopoff, 1996). To demonstrate the influence of
a cutoff, we restrict the previous case to
$$\begin{array}{lllll}
p(t) & = {1 \over 1 - {({t_{max} \over t_0})}^{-\mu}}~~{\mu \over t_0}
\biggl({t \over t_0}\biggl)^{-(1 + \mu)} & , & ~~~~t_0 \leq t \leq
t_{max}~ \\      & = 0 & ,  &~~~~ 0 < t <  t_0  & ,
\end{array}
\eqno(A6)$$
which is normalized.  Substitution in eq. (2) yields
$$ P(t') =  {\mu \over (t+t')^{1+\mu}}~{1 \over t_{}^{-\mu} -
t_{max}^{-\mu}}~~~~~~~~~~~~.  \eqno(A7) $$
For $t \ll t_{max}$, the second factor of (A7) is $t^\mu,$ which is
the same as letting $t_{max} \rightarrow \infty .$ Thus we recover the
previous case of the simple power law without truncation.
 
The interesting regime is found when $t$ is not very small compared to
$t_{max}$. Consider the case ${t \rightarrow t_{max}}$. From eq. (A7)
we get
$$ P(t') \simeq {1 \over t_{max} - t}~~~~~~~~~~~~~~~~, \eqno(A8) $$
which is independent of $t'$ and becomes very large as $t \to
t_{max}.$  This case is identical to that of the uniform distribution
(A1) above.  Thus it is not unexpected that the longer we wait, the
shorter will be the expected time until the next event. Without
truncation, the result is reversed. There is a crossover between the
truncated and untruncated cases as illustrated in figure 6. In the
figure, we take $\mu = 3$ as in figure 5a, $t_{max}=100 t_0$ and show
$P(t')$ for $t=10 t_0$, $90 t_0$ and $98 t_0$. For $t =10 t_0$,
$P(t')$ is found to be much larger than $p(t')$ in the tail, as in the
previous untruncated case. For $t=90 t_0$, $P(t')$ is defined only for
$0\leq t' \leq 10 t_0$. In agreement with (A7), we see that $P(t')$
becomes almost constant and close to ${1 \over 10 t_0}$. For $t=98
t_0$, $P(t')$ is defined only for $0\leq t' \leq 2 t_0$ and is close
to ${1 \over 2 t_0}$. This illustrates the crossover from a longer
expected time when $t$ is small, to a shorter expected time as $t$
approaches $t_{max}$.
 
Since the truncated power-law distribution is very close to a uniform
distribution for times near $t_{max},$ we expect that the truncated
lognormal distribution and truncated Weibull distributions with $\mu <
1$  will also have a shorter time until the next event, the longer we
have been waiting for an earthquake to happen; thus  we expect a
crossover in the response to Q. between these truncated and
untruncated cases as well.

\newpage
\begin {center}
\centering {\large\bf Acknowledgments}
\end {center}
\vspace {0.25cm}
 
This research has been partially supported by the NSF/CNRS under the
US/France International Cooperation program. This paper is Publication
no. 4669 of the Institute of Geophysics and Planetary Physics,
University of California, Los Angeles, and Publication no. YYYY of the
Southern California Earthquake Center.

 \pagebreak
\vspace*{0.3cm}
\begin{center}
\centering{TABLE 1\,:  RESPONSE TO Q.}
\end{center}
 
\begin{table*}[h]
\begin{center}
\begin{tabular}{|l|c|c|c|c|} \hline
Distribution &  Short times & Long Times \\ \hline
Exponential & 0 & 0\\ \hline
Periodic & -- &  \\ \hline
Uniform & -- &   \\ \hline
Gaussian &   & -- \\ \hline
Semigaussian & -- &  -- \\ \hline
Lognormal & -- & +  \\ \hline
~~~~~~${1\over 4}{\sqrt x}e^{-\sqrt x}$ & -- &  + \\ \hline
Weibull $(m > 1)$ & -- & --  \\ \hline
~~~~~~Rayleigh (m=2) & -- & --  \\ \hline
Weibull $(m < 1)$ & + & +  \\ \hline
Power Law & -- & +  \\ \hline
Truncated Power Law  & + & --  \\ \hline
\end{tabular}
\end{center}
\end{table*}

\newpage
\begin {center}
\centering {\large\bf Figure Captions}
\end {center}
\vspace {0.25cm}
 
Figure 1. Expected time of the next earthquake as a function of the
elapsed time since the last one.  The response to Q. is given by the
position of the curve with respect to $1$.
 
\vspace{0.5cm}
Figure 2. Uniform distribution: $p(t')$ and $P(t')$ for $t = (4/3) t_0$.
 
\vskip 0.5cm
Figure 3. Lognormal distribution with $\sigma = t_0$: $p(t')$ and $P(t')$
for $t=2t_0$ and $t=5 t_0$. a) short times\,; b) long times.

\vskip 0.5cm
Figure 4. Weibull distribution.
 
a) $m = 2$ (Rayleigh distribution), corresponding to a tail decaying
faster than an exponential. $P(t')$ is shown for $t=t_0$ and $t=2t_0$
together with $p(t')$.
 
b) $m=1/2$ corresponding to a  tail decaying slower than an
exponential. $P(t')$ is shown for $t=t_0$, $t=2 t_0$ and $t=10 t_0$,
together with $p(t')$.
 
\vskip 0.5cm
Figure 5. Power-law distribution.
 
a) $\mu = 3$; $p(t)$ possesses a finite mean and variance.
$P(t')$ is shown for $t=10 t_0$ and $t=100 t_0$ together with $p(t')$.
 
b) $\mu = 1$; the mean and variance are not defined.
$P(t')$ is shown for $t=10 t_0$ and $t=100 t_0$ together with $p(t')$.
 
\vskip 0.5cm
Figure 6. Truncated power-law distribution with $\mu = 3$ and $t_{max}
= 100 t_0$. $P(t')$ is shown for $t=10 t_0$, $90 t_0$ and $98 t_0$.

\vspace {1.0in}
 
\noindent
{\it Department of Earth and Space Science and Institute of Geophysics
and Planetary Physics\\ University of California, Los Angeles,
California 90095\\
and Laboratoire de
Physique de la Mati\`ere Condens\'ee, CNRS URA190\\ Universit\'e des
Sciences, B.P. 70, Parc Valrose, 06108 Nice Cedex 2, France\\
\indent (D.S.)}
 
\vspace {0.5cm}
 
\noindent
{\it Department of Physics and Institute of Geophysics and Planetary
Physics\\ University of California, Los Angeles, California 90095\\
\indent (L.K.)}
 
\end{document}